\documentclass[twocolumn,showpacs,showkeys,pra,aps]{revtex4-1}

\usepackage[dvipsnames]{xcolor}
\usepackage{tikz}
\usetikzlibrary{chains,backgrounds,scopes,decorations.pathmorphing}
\usetikzlibrary{shapes.gates.logic.US,shapes}
\usetikzlibrary{arrows,patterns}
\usetikzlibrary{matrix}

 \usepackage{bbm}
 \usepackage{amsmath}
 \usepackage{amssymb}
 \usepackage{latexsym}
 \usepackage{amsfonts}
 \usepackage[caption=false]{subfig}
 \usepackage{epsfig}
 \usepackage{psfrag}
 \usepackage{color}
 \definecolor{darkblue}{rgb}{0,0,.5}
 \usepackage[linktocpage, colorlinks=true ,linkcolor=darkblue, citecolor=darkblue]{hyperref}
 \usepackage[all]{hypcap}

 \newcommand{\ket}[1]{\left|#1\right>}
 \newcommand{\bra}[1]{\left<#1\right|}
 \newcommand{\expval}[1]{\left< #1 \right>}

 \newcommand{\nn}{\nonumber\\}
 
 \newcommand{\f}[1]{\mbox{\boldmath$#1$}}

 \newcommand{\bea}{\begin{eqnarray}}
 \newcommand{\ea}{\end{eqnarray}}
 \newcommand{\eea}{\end{eqnarray}}
 
 \newcommand{\trace}[1]{{\rm Tr}\left\{ #1 \right\}}
 
 \newcommand{\traceS}[1]{{\rm Tr_S}\left\{ #1 \right\}}

 \newcommand{\dt}{\frac{\rm d}{\rm dt}}
 \newcommand{\cu}[1]{\langle\langle #1 \rangle\rangle}
 \newcommand{\op}[1]{{\mathbf X}_{#1}}
 \newcommand{\ii}{{\rm i}}
\begin{document}

\title{Equation of motion method for Full Counting Statistics: Steady state superradiance}

\author{Malte Vogl$^*$, Gernot Schaller, Eckehard Sch\"oll and Tobias Brandes}

\affiliation{Institut f\"ur Theoretische Physik, Technische Universit\"at Berlin, Hardenbergstr. 36, 10623 Berlin, Germany}

\begin{abstract}
  For the multi-mode Dicke model in a transport setting that exhibits collective boson transmissions, we construct the equation 
of motion for the cumulant-generating function. Approximating the exact system of equations at 
the level of cumulant-generating function and system operators at lowest order, allows us
to recover master equation results of the Full Counting Statistics for certain parameter regimes at very low cost of computation. 
The thermodynamic limit, that is not accessible with the master equation approach, can be derived analytically for different 
approximations. 
\end{abstract}



\maketitle
\section{Introduction}
The fluctuations of particles tunneling randomly through a quantum structure are sensitive indicators for
particle interactions~\cite{blanter2000}, coherences~\cite{ferrini2009} or collective effects~\cite{gabelli2009}. 
Originating from the field of quantum optics~\cite{cook1981}, the theory of Full Counting Statistics (FCS) has 
proven to be a versatile tool for the evaluation of these fluctuations, that can be conveniently described by the 
cumulants of the underlying stochastic process~\cite{schloegl1983}.
FCS is successfully applied in electronic systems~\cite{levitov1996,bagrets2003}, where the charge fluctuations can be measured
by a nearby quantum point contact~\cite{flindt2009}.
While for various small-sized quantum systems the FCS can be derived analytically~\cite{andreev2000}, 
the calculation of the full statistics, i.e., all higher cumulants, of the emitted particles for 
an 
arbitrarily 
large system size is cumbersome
due to the required diagonalization of the large Liouvillian ${\cal L}(\chi)$, that describes the system dynamics.

Apart from numerical methods the most obvious approach to this problem is to make further approximations concerning 
the factorization of higher order correlation functions, as is routinely done in, e.g., the Hartree-Fock approach in solid
state physics~\cite{richter2009}. While this factorization is usually applied at the level of specific observables, 
e.g., for two-electron processes, in this work we propose a factorization for the correlations of the 
cumulant-generating function (CGF) and system operators derived within a weak-coupling theory.
One of the most basic and yet size-scalable models is the Dicke model~\cite{dicke1954} for 
super-radiant decay of an initially excited atomic cloud interacting with a radiation field~\cite{gross1982}. 
In the limit of small sample size, the dynamics of the system can be analyzed in the 
symmetrical angular momentum basis~\cite{bonifacio1971} and analytic results for higher correlations 
of the transient pulse of emitted bosons can be obtained~\cite{haake1972}. A semi-classical propagator for transient
Dicke superradiance was derived in~\cite{braun1998}.

In this work, we consider an extended multi-mode Dicke model in a transport setup~\cite{vogl2011}, in a limit where back-action between 
counted particles and the system state can be neglected. Contrary to previous work we have therefore a steady-state transport setting 
that allows us to obtain analytic long-time results. 
For this aim, we construct an equation of motion (EoM) for the CGF, the solution of which requires us
to make factorization assumptions for correlations of counting and system operators. 
We show that in certain limits even factorization at lowest level allows one to retrieve the FCS for arbitrary system size 
and access the thermodynamic limit. The semi-classical results for transient effects 
can be obtained as limiting cases~\cite{brandes2005}. 

With the recent progress in single-photon detectors~\cite{hadfield2009a} and the rising interest in collective boson transport,
e.g., in thermal transport~\cite{deliberato2011} or new types of lasers, like phonon and superradiant lasers~\cite{grudinin2010,bohnet2012}, 
we expect our findings to be relevant for a broader community.

This work is structured as follows. In Sec.~\ref{sec:model} we introduce the model under consideration
and briefly introduce the underlying theory. In Sec.~\ref{sec:exactsol} we present exact EoM for the CGF
and correlations. In Sec.~\ref{sec:mgf_eom} we consider arbitrary system size $N$ 
and compare 
the solutions for different approximations that close the EoM (Sec.~\ref{sec:jz_eom}). We then derive analytic expressions for 
the thermodynamic limit of the FCS (Sec.~\ref{sec:thermodynamic}), examine the scaling behavior of the cumulants (Sec.~\ref{sec:scaling}),
and discuss the quality of the different approximations (Sec.~\ref{sec:quality}). 
We finish in Sec.~\ref{sec:concl} with a conclusion.
\section{Model}\label{sec:model}
Our model consists of two bosonic reservoirs, source $S$ and drain $D$, that are coupled collectively by a medium of $N$ two-level systems with 
identical level splitting $\Omega$. It is described by an extended multi-mode Dicke Hamiltonian~\cite{vogl2011}
\bea
H &=& \frac{\Omega}{2} J_z + \sum_{\alpha =S,D} \sum_{k} \omega_{k\alpha}b_{k\alpha}^\dagger b_{k\alpha}\nn
&&+ J_x \sum_{\alpha =S,D} \sum_{k} \left(h_{k\alpha} b_{k\alpha}^\dagger + h_{k\alpha}^* b_{k\alpha} \right)
\eea 
with collective spin operators $J_{x,z} \equiv \sum_{i=1}^N \sigma_i^{x,z}$, where 
the operator $b_{k\alpha}^\dagger$ creates a boson 
of frequency $\omega_{k\alpha}$ in reservoir $\alpha$ with the corresponding coupling constant $h_{k\alpha}$. 

We obtain the counting statistics of the total number $n$ of bosons exchanged between the medium and the drain 
by formally introducing~\cite{schaller2009} a bookkeeping operator
$d^\dagger = \sum_{n=-\infty}^\infty \ket{n+1}\bra{n}$ in the Hamiltonian via 
$b^\dagger_{k D} \to b^\dagger_{kD} \otimes d^\dagger$ and similarly 
for the annihilation operator. 
The bookkeeping operator $d^\dagger$ increases the occupation
of a virtual detector by one unit for every boson created in the drain, independent on the mode $k$.
\begin{figure}[t]
\includegraphics[width=0.3\textwidth,clip=true]{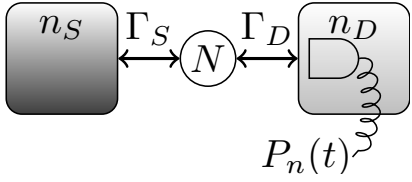}
\caption{Sketch of the model: Bosonic excitations are absorbed/emitted with rates $\Gamma_{S/D}$ in and out of the medium of $N$ 
two-level systems to source or drain, respectively (occupations $n_S > n_D$). 
The number of total boson emissions into the drain is counted to construct the distribution $P_n(t)$, denoting the 
probability for $n$ particles effectively emitted into the drain after time $t$.}
\label{fig:sketch}
\end{figure}
In the following, we consider the $N$-two-level 
medium and the bookkeeping operator as the system and assume a weak coupling between system 
and reservoirs. This allows us to derive a Lindblad master equation in Born, Markov, and secular 
approximation, where we assume initial decoupling of system and bath, a memory-less bath and neglect fast rotating terms 
in comparison with the system time-scale~\cite{breuerpetruccione2002}. 

With the extension of the Hamiltonian by the bookkeeping operator, 
the system state $\rho^{(n)} \equiv \bra{n} \rho \ket{n}$ is conditioned on the number of bosons $n$ measured in the detector. 
The master equation assumes the form 
\bea\label{eq:nresdensity}
\dot{\rho}^{(n)}(t) = {\cal L}_0 \rho^{(n)} + {\cal L}_+ \rho^{(n-1)} + {\cal L}_- \rho^{(n+1)}\,,
\eea 
where $\cal{L}_\pm$ describe emissions out of (+) and into (--) the system to and from the drain, while ${\cal L}_0$ 
describes the remaining evolution. Explicitly, the super-operators are given by~\cite{vogl2011} 
\bea\label{eq:nresolved}
{\cal L}_{0} \rho &=& -\frac{\ii \Omega}{2} \left[J_z,\rho \right] -\sum_{\alpha=S,D} \frac{\Gamma_\alpha n_\alpha}{2} 
\left\{J_- J_+,\rho \right\}\nn 
&&-\sum_{\alpha=S,D} \frac{\Gamma_\alpha (1+n_\alpha)}{2} \left\{J_+ J_- , \rho \right\}\nn
&&+ \Gamma_S \left[1+n_S\right] J_- \rho J_+ + \Gamma_S n_S J_+ \rho J_-\,,\nn
\cal{L}_+ \rho &=& \Gamma_D \left[1+n_D\right] J_- \rho J_+\,,\nn
\cal{L}_- \rho &=& \Gamma_D n_D J_+ \rho J_-\,,
\eea
where \mbox{$\Gamma_\alpha \equiv 2 \pi \sum_k |h_{k\alpha}|^2 \delta(\Omega - \omega_{k\alpha})$} denotes the spontaneous boson 
emission rate of a single two-level system into a vacuum 
reservoir $\alpha$ and $\left\{\cdot,\cdot \right\}$ is the anticommutator.
The numbers $n_\alpha$ denote the stationary occupation of bath modes at system transition frequency $\Omega$:
For the case of thermal baths one has, e.g., \mbox{$n_\alpha = \left(e^{\beta_\alpha \Omega}-1\right)^{-1}$}, where $\beta_\alpha$
is the inverse temperature of bath $\alpha$.

The probability of counting $n$ bosons after time $t$ is given by $P_{n}(t) = \trace{\rho^{(n)}(t)}$. 
The $n$-resolved master equation~\eqref{eq:nresdensity} can be Fourier-transformed by introducing a counting field 
$\chi$ via $\rho(\chi,t) \equiv \sum_{n} \rho^{(n)}(t)e^{\ii n \chi}$ leading to a generalized master equation for collective spontaneous 
emission and absorption 
 $\dt \rho(\chi,t) = {\cal L}(\chi)\rho(\chi,t)$ with ${\cal L}(\chi) = {\cal L}_0 +e^{+\ii \chi}{\cal L}_+ +e^{-\ii \chi}{\cal L}_-$. 
In case of a single reservoir and neglect of the counting field, this reproduces previous master equations~\cite{agarwal1974}.
 

The FCS of the bosonic detection probability $P_{n}(t)$ for $n$ bosons after time $t$ 
counted at the drain reservoir is given by the cumulants $\cu{n_k(t)}\equiv (-\ii \partial_\chi)^k {\cal C}(\chi,t)|_{\chi =0}$,
denoted by double brackets,  
where 
\bea
{\cal C}(\chi,t) \equiv {\rm ln} \trace{\rho(\chi,t)} = {\rm ln} \trace{e^{{\cal L}(\chi)t} \rho_0}
\eea
is the CGF, cf. e.g.,~\cite{flindt2008}. In the stationary limit
the CGF can be approximated by the eigenvalue $\lambda_0(\chi)$ with smallest modulus, fulfilling $\lambda_0(0)=0$,
such that \mbox{$\lim\limits_{t\to\infty}{\cal C} (\chi,t)\approx \lambda_0(\chi) t$}. 
Throughout this work we are investigating the long-time current cumulants
$\cu{I_k} \equiv \lim_{t\to\infty}\dt \cu{n_k(t)}$. 

Calculations are conveniently performed in the angular momentum basis, where $j=\frac{N}{2}$ and
\bea
J_z \ket{j,m} &=& 2 m \ket{j,m}\,,\nn
J_\pm \ket{j,m} &=& \sqrt{j(j+1)-m(m \pm 1)}\ket{j,m \pm 1}\,,\nn
\f{J}^2 \ket{j,m} &=& 4 j(j+1) \ket{j,m}\,,  
\eea
with $\f{J}=(J_x,J_y,J_z)^T$, and $\left[J_+, J_- \right] =J_z$. 
In our model the system size $N$ is fixed, so that we can treat the expectation value 
\mbox{$\expval{\f{J}^2}= \f{J}^2 = 2 N (\frac{N}{2}+1)$} as a constant. For later use note that 
\mbox{$J_- J_+ = \frac{1}{4}\left[\f{J}^2-J_z^2\right]-\frac{1}{2} J_z$} and \mbox{$\left[J_z,J_\pm \right]=\pm 2 J_\pm$}.
\section{Exact results}\label{sec:exactsol}
For the stationary properties of the expectation values $\expval{J_z^\alpha}$ we have an exact benchmark:
Even for the case of non-equilibrated reservoirs the stationary state of the system, defined by ${\cal L}(0)\bar{\rho}=0$, is a thermal one, 
$\bar{\rho} \propto e^{-\frac{\bar{\beta} \Omega}{2} J_z}$, with a temperature $\bar{\beta}^{-1}$ that corresponds to coupling
the system to a single fictitious reservoir with the weighted average occupation 
\bea\label{eq:efftherm}
\bar{n} = \frac{\Gamma_S n_S + \Gamma_D n_D}{\Gamma_S + \Gamma_D} \equiv 
\frac{1}{e^{\bar{\beta} \Omega}-1}\,.
\eea
Such simple average occupations exist also for an arbitrary number of
baths whenever the operator structures of the couplings are identical~\cite{schaller2011}.
Since in addition our system only supports a single allowed transition
frequency, it effectively thermalizes at some average temperature even
when coupled to baths at different temperatures.
%

Using this effective thermalization we find the ratio of the matrix elements $\bar{\rho}_m =\bra{jm}\bar{\rho}\ket{jm}$ 
of the tri-diagonal Liouvillian $\frac{\bar{\rho}_{m+1}}{\bar{\rho}_m}=\frac{\bar{n}}{\bar{n}+1}$. 
In connection with normalization $\sum_m \bar{\rho}_m = 1$, this implies that the steady state expectation values
\bea
\bar{\expval{J_z^k}}_{ex} &=& \trace{J_z^k \bar{\rho}}\label{eq:exactJz2}\,,
\eea
with $k=1,\,2,\,\dots$ can be calculated exactly (not shown for brevity) for 
arbitrary system sizes $N$.

The evolution of the CGF for the emitted bosons at the drain can be calculated using
\mbox{${\cal C}(\chi,t)={\rm ln} \expval{e^{\ii n \chi}} = {\rm ln} \sum_n e^{\ii n \chi} P_n(t)$}, which yields
\bea
\label{eq:dotmgffull}
 \dt{\cal C}(\chi,t) &=&  \frac{\sum_n e^{\ii n \chi}\trace{\dot{\rho}^{(n)}}}{\expval{e^{\ii n \chi}}}\\
&=&n_D \left(f[\chi]-f[-\chi]\right)  \frac{\expval{e^{\ii n \chi} J_z}}{\expval{e^{\ii n \chi}}} \nn
&&+\frac{n_D}{2}\left(f[\chi] + f[-\chi] \right)\left[\f{J}^2 - \frac{\expval{e^{\ii n \chi} J_z^2}}{\expval{e^{\ii n \chi}}}\right] \nn
&&+\frac{f[\chi]}{2} \left[\f{J}^2 -\frac{\expval{e^{\ii n \chi} J_z^2}}{\expval{e^{\ii n \chi}}} 
+ 2 \frac{\expval{e^{\ii n \chi} J_z}}{\expval{e^{\ii n \chi}}} \right]\,,\nonumber
\eea
where 
\bea\label{eq:fchifct}
f[\chi]\equiv \frac{1}{2} \Gamma_D (e^{\ii \chi}-1)
\eea 
contains the counting field. Thus, the time-evolution of the CGF couples only to 
two other correlations $\expval{e^{\ii n \chi} J_z}$ and $\expval{e^{\ii n \chi} J_z^2}$. 
We therefore calculate the EoM for correlations of arbitrary order $\alpha \geq 1$, which is given by 
\bea\label{eq:dotcorrelations}
&&\dt \expval{e^{\ii n \chi} J_z^\alpha}\equiv \sum_n \trace{e^{\ii n \chi} J_z^\alpha \dot{\rho}^{(n)}(t)}\\
 &=&\biggl<e^{\ii n \chi}\biggl[ \Gamma_S n_S \left[\left(J_z + 2 \cdot\mathbbm{1}\right)^\alpha -J_z^\alpha \right]\op{-}\nn
&&+\Gamma_S(1+n_S) \left[\left(J_z - 2 \cdot\mathbbm{1}\right)^\alpha -J_z^\alpha \right]\op{+}\nn
&&+\Gamma_D n_D \left[e^{-\ii \chi}\left(J_z + 2 \cdot\mathbbm{1}\right)^\alpha -J_z^\alpha \right]\op{-}\nn
&&+\Gamma_D (1+n_D)\left[e^{\ii \chi}\left(J_z - 2 \cdot\mathbbm{1}\right)^\alpha -J_z^\alpha \right]\op{+}\biggr]\biggl>\,,\nonumber
\eea
with
\bea\label{eq:xpmfct}
\op{\pm}=\frac{1}{4}\left(\f{J}^2 - J_z^2 \pm 2 J_z\right)\,,
\eea
and thus couples the evolution of $\expval{e^{\ii n \chi} J_z^\alpha}$ to all powers of 
$\expval{e^{\ii n \chi} J_z^{\alpha'}}$ where $\alpha'\in\{0,1,\ldots,\alpha+2\}$. 
This 
leads to a hierarchy problem, such that the straightforward solution of these EoMs without factorizations is 
difficult for large $j=\frac{N}{2}$. Equations~\eqref{eq:dotmgffull}~--~\eqref{eq:xpmfct} are the first central result of this paper.    

For $N=1$ (i.e., $j=\frac{1}{2}$) however, 
the above 
system can be solved exactly. For this special case 
we have $J_z = \sigma_z$ and $\sigma_z^2={\mathbbm 1}$ and thus we can solve the system of equations
without the need of any factorization. We introduce the abbreviation 
$A(\chi,t)\equiv \frac{\expval{e^{\ii n \chi} J_z}}{\expval{e^{\ii n \chi}}}$ and obtain
\bea\label{eq:N=1}
  \dt{\cal C}(\chi,t) &=& -\left(e^{-\ii \chi}-1\right)\frac{\Gamma_D}{2} n_D\left[A(\chi,t)-1\right]\nn
&&+\left(e^{+\ii \chi}-1\right)\frac{\Gamma_D}{2}\left[1+n_D\right]\left[A(\chi,t)+1\right]\nn
\dt A(\chi,t)&=& \underbrace{\frac{\dt \expval{e^{\ii n \chi} J_z}}{\expval{e^{\ii n \chi}}}}_I-A(\chi,t)\left(\dt{\cal C}(\chi,t) \right)\,,
\eea  
with
\bea
 I &=& -\Gamma_S n_S \left[A(\chi,t)-1\right] -\Gamma_S \left[1+n_S \right] \left[A(\chi,t)+1\right]\nn
 &&-\frac{\Gamma_D}{2} n_D \left(e^{-\ii \chi}+1 \right)\left[A(\chi,t)-1 \right]\nn
 &&-\frac{\Gamma_D}{2} \left[1+n_D \right] \left(e^{+\ii \chi}+1 \right)\left[A(\chi,t)+1 \right]\,.
\eea
The quadratic equation for $A(\chi,t)$, cf. Eq.~\eqref{eq:N=1}, can be solved and inserted in the equation for the CGF. Since we
are interested in long-time dynamics, we can directly take derivatives with respect to~$\chi$ and send $t \to \infty$ to obtain
the corresponding stationary cumulants. Note that to circumvent the quadratic equation one could alternatively calculate the 
dynamics of the moment-generating function ${\cal M}(\chi,t) \equiv \expval{e^{\ii n \chi}}$, where one has to solve a
system of coupled linear equations and then construct the cumulants from the moments.
As expected, the thus obtained CGF coincides in the long-time limit with the eigenvalue $\lambda_0(\chi)$ with smallest modulus, obtained by diagonalizing the 
corresponding $2\times 2$ 
Liouvillian, and is given by
\bea
&&{\cal C}^{N=1}(\chi,t) = -\frac{1}{2} t\biggl[\Gamma_S (1+2 n_S)+\Gamma_D (1+2 n_D)\\
&-&\biggl\{\Gamma_S \Gamma_D \left[ e^{-\ii \chi}(1+n_S)n_D+e^{+\ii \chi} n_S(1+n_D)\right]\nn
&+& \frac{1}{4}\left[2\Gamma_S \Gamma_D +(\Gamma_S (1+2 n_S))^2+(\Gamma_D (1+2 n_D)) \right]\biggr\}^{\frac{1}{2}} \biggr]\,.\nonumber
\eea

In contrast, for large $N$ it is much more difficult to obtain the CGF, as already the size of the Liouvillian grows as $(N+1)\times(N+1)$.
However, for the long-time limit of the first cumulant $\cu{I_1}$ we can make use of Eq.~\eqref{eq:exactJz2} and the fact 
that trace conservation implies $\trace{{\cal L}(0) \rho}=0$, such that we find
\bea\label{eq:n1me}
\cu{I_1^{\rm ME}} &=& -\ii \traceS{{\cal L}'(0) \bar\rho}\nn 
              &=& \left(n_S-n_D\right) \frac{\Gamma_S \Gamma_D}{\Gamma_S + \Gamma_D} \sigma_N\,,\\
      \sigma_N&=&\frac{\left(N - 2 \bar{n}\right) \left(1 + \bar{n}\right)^{N+1} 
                  + \bar{n}^{N+1}\left(2 + N + 2 \bar{n}\right)}{\left(1+\bar{n}\right)^{N+1} - \bar{n}^{N+1}}\,,\nonumber
\eea
where the superscript ${\rm ME}$ denotes the master equation solution. This result (which we obtained previously~\cite{vogl2011})
gives us one analytic benchmark for approximate solutions. 

Due to $\left[{\cal L}'(\chi),{\cal L}(\chi) \right] \neq 0$, the above approach yields no simple analytic results 
for higher cumulants (that require higher derivatives with respect to~$\chi$). The corresponding expressions still involve
the full $\chi$-dependent Liouvillian, that needs to be diagonalized for a solution. 
To bypass the expensive numerical calculation, which becomes unfeasible already for moderate $N$,
we have to apply additional approximations. 
\section{Approximate Equations of Motion}\label{sec:mgf_eom}
Eq.~\eqref{eq:dotmgffull} and \eqref{eq:dotcorrelations} constitute the EoM for the full CGF. To circumvent the 
problem of the hierarchy of operator equations, we introduce a lowest-order factorization at the level of 
the CGF and system operators, and show that already this very crude assumption gives useful results
comparable to the master equation solution. 

The simplest way to make Eq.~\eqref{eq:dotmgffull} solvable is to add one further 
assumption to the Born-Markov-Secular approximation,
namely that the counting operator $n$ has no correlations in any order $\alpha$ with the system operators
\bea\label{eq:factoring_assumption}
\expval{e^{\ii n \chi} J_z^\alpha} \approx \expval{e^{\ii n \chi}}\expval{J_z^\alpha}\,.
\eea
This assumption is strictly valid only when the statistics of particles counted at the drain and the system state are independent.
Inserting the approximation 
Eq.~\eqref{eq:factoring_assumption} into Eq.~\eqref{eq:dotmgffull} yields another central result of this work: 
an explicit equation for the CGF
\bea\label{eq:dotmgf}
  \dt {\cal C}(\chi,t) &=& n_D \left(f[\chi]-f[-\chi]\right)  \expval{J_z} \nn
&&+\frac{n_D}{2}\left(f[\chi] + f[-\chi] \right)\left[\f{J}^2 -\expval{J_z^2}\right] \nn 
&&+\frac{f[\chi]}{2} \left[\f{J}^2 -\expval{J_z^2} + 2 \expval{J_z} \right]\,,
\eea
with $f[\chi]$ given by Eq.~\eqref{eq:fchifct}.

Independent of the choice of solution for $\expval{J_z^{(1,2)}}$, taking derivatives with respect to $\chi$ shows 
that all time derivatives of odd ($\cu{n_{2 k+1}}$) and even ($\cu{n_{2 k}}$) cumulants are identical, respectively, and are given by
\bea
\dt \frac{\cu{n_{2k+1}}}{\Gamma_D} &=&  n_D \expval{J_z} + \frac{1}{4} \left[\f{J}^2 -\expval{J_z^2}+2 \expval{J_z} \right]\,,\\
\dt \frac{\cu{n_{2k}}}{\Gamma_D}   &=& \frac{1}{2} \left[\left(n_D + \frac{1}{2}\right) \left(\f{J}^2-\expval{J_z^2} \right) +2 \expval{J_z} \right]\,.\nonumber
\eea 
Therefore, provided that Eq.~\eqref{eq:factoring_assumption} is approximately valid, the statistics of the model can be retrieved 
with knowledge of only $\expval{J_z}$ and $\expval{J_z^2}$.
\subsection{Approximate solutions for large spin}\label{sec:jz_eom}
For our specific system we are able to obtain exact steady state expressions for the expectation values of the system
operator for arbitrary power $\expval{J_z^\alpha}$, cf. Eq.~\eqref{eq:exactJz2}.
If one applied our approach to other systems, this would generally not be the case. 
To close the system of equations we would therefore have 
to calculate the EoM for the system operators as well, even when factoring correlations of the type of 
Eq.~\eqref{eq:dotcorrelations}. 

To elucidate the validity of the central approximation Eq.~\eqref{eq:factoring_assumption}, we
also calculate solutions of Eq.~\eqref{eq:dotmgf} using approximate expressions for the evolution of the system operators. 
For this aim, using Eq.~\eqref{eq:nresolved} yields an EoM for powers of $J_z$ given by
\bea\label{eq:Jzalpha}
\dt \frac{\expval{J_z^\alpha}}{\Gamma} &=& \bar{n} 
\expval{\left[\left(J_z + 2 \cdot\mathbbm{1}\right)^\alpha -J_z^\alpha \right]\op{-}}\nn
&&+(\bar{n}+1)\expval{\left[\left(J_z - 2 \cdot\mathbbm{1}\right)^\alpha -J_z^\alpha \right]\op{+}}\,,
\eea
where $\Gamma = \Gamma_S + \Gamma_D$, $\op{\pm}$ are given in Eq.~\eqref{eq:xpmfct} and $\bar{n}$ is given in Eq.~\eqref{eq:efftherm}.  
Since $\op{\pm}\propto J_z^2, J_z$, the evolution of an operator $J_z^\alpha$ couples to all operators 
$J_z^{\alpha'}$ with $\alpha' \in \left\{1,2,\ldots,\alpha+1 \right\}$. 
The system of coupled linear differential equations for
 $\expval{J_z^\alpha}$ gives again rise to a hierarchy problem, now on the level of system operators, 
and can be solved approximately for large $j$. 
Factoring the expectation value at any level (e.g. 
$\expval{J_z^{\alpha+\alpha'}}\approx\expval{J_z^\alpha}\expval{J_z^{\alpha'}}$) 
will lead to a coupled set of non-linear 
differential equation, that in the time-dependent case requires numerical solutions. In our approach we are interested in
steady-state results and will therefore always use long-time expectation values.
%
\subsubsection*{Approximation 1}
To estimate the errors introduced by the factorizations on top of the general one~\eqref{eq:factoring_assumption},
we can use solutions of Eq.~\eqref{eq:dotmgf} as a benchmark, where the exact stationary results $\expval{\bar{J_z^1}}_{ex}$ 
and $\expval{\bar{J_z^2}}_{ex}$ of Eq.~\eqref{eq:exactJz2} for $\alpha = 1,\,2$ are inserted.
Since we have applied our general factorization Eq.~\eqref{eq:factoring_assumption} even when using
the exact expressions we will denote this solution as Approximation~1. 
In the following we differentiate between two further levels of factorization.
Higher order approximations can easily be constructed.
\subsubsection*{Approximation 2}
Here, in addition to~\eqref{eq:factoring_assumption} we take only the dynamics of $\expval{J_z}$ into account, 
assume $\expval{J_z^2} \approx \expval{J_z}^2$ 
in Eq.~\eqref{eq:Jzalpha} with $\alpha=1$ (Approximation~2).
This is valid whenever the variance vanishes and leads to a single quadratic equation,
\bea
\dt \frac{\expval{J_z}}{\Gamma} &=& -\left(2 \bar{n}+1 \right)\expval{J_z} +\frac{1}{2}\expval{J_z}^2 - \frac{1}{2} \f{J}^2\,,
\eea 
which can be solved analytically for all times.
Since we are interested in steady state dynamics of the FCS, we take the long-time limit and insert 
the solution in Eq.~\eqref{eq:dotmgf} where we again assume $\expval{J_z^2} \approx \expval{J_z}^2$.
\subsubsection*{Approximation 3}
Alternatively, again in addition to~\eqref{eq:factoring_assumption} we take $\expval{J_z^2}$ in Eq.~\eqref{eq:Jzalpha} for \mbox{$\alpha=1$} into account and factorize third order correlations
 \mbox{$\expval{J_z^3} \approx \expval{J_z}\expval{J_z^2}$} for Eq.~\eqref{eq:Jzalpha} with $\alpha=2$ (Approximation~3).
This yields a coupled system of (non-linear) differential equations,
\bea
\dt \frac{\expval{J_z}}{\Gamma} &=& -\left(2 \bar{n}+1 \right)\expval{J_z} +\frac{1}{2}\expval{J_z^2} - \frac{1}{2} \f{J}^2\,,\\
\dt \frac{\expval{J^2_z}}{\Gamma} &=& \left(2 \bar{n}+1\right)\left[\f{J}^2 -3 \expval{J_z^2}\right]-\left[\f{J}^2-2\right]\expval{J_z}\nn
&&+\expval{J_z^2}\expval{J_z}\,,\nonumber
\eea
 that in general has to be solved numerically.
However, the steady state solution can be obtained analytically and again is inserted into Eq.~\eqref{eq:dotmgf}, 
where apart from Eq.~\eqref{eq:factoring_assumption} we now do not have to assume further factorizations.
\subsection{Thermodynamic limit for the cumulant-generating function}\label{sec:thermodynamic}
The full CGF for the master equation solution is not accessible for arbitrary system size, 
since the diagonalization of the large Liouvillian is computationally cumbersome. 
For the approximate case however, solutions can be easily derived by 
solving Eq.~\eqref{eq:dotmgf} with the different approximations and we are thus able to examine the thermodynamic limit of 
infinite system size. 

The obtained expressions for the approximate CGFs, with $\alpha$ marking the applied approximation, are
\mbox{${\cal C}^{\alpha}(\chi,t)=\mathbf{F}(\chi,t) \tilde{{\cal C}}^{\alpha}$} with
\bea\label{eq:appcgf}
\tilde{\cal C}^{1}&=& \frac{\left(\bar{n}+1\right) \left(N-2 \bar{n} \right)+\bar{n} \left(\frac{\bar{n}}{1+\bar{n}} \right)^N \left(N+2 \bar{n}+2 \right)}
{\bar{n}\left[1-\left(\frac{\bar{n}}{1+\bar{n}}\right)^N\right]+1}\,,\nn
\tilde{\cal C}^{2}&=&-\left(2 \bar{n}+1\right)+\sqrt{(2\bar{n}+1)^2+N (2+N)} \,,\nn
\tilde{\cal C}^{3}&=&-\frac{1}{2}\left(3 + 6\bar{n}-(2\bar{n}+1)^{-1}\right)\nn
&&+\sqrt{\frac{1}{4}\left(3+6\bar{n}-(2\bar{n}+1)^{-1}\right)^2 +N(2+N)}\,,
\eea
where the common function 
\bea\label{eq:commonfct}
\mathbf{F}(\chi,t) &=& \Gamma_D t \biggl[(e^{\ii \chi}-1)(n_D+1)\bar{n}\nn
&&+(e^{-\ii \chi} -1)n_D (\bar{n}+1) \biggr]\,,
\eea
is shared by all approximations. The form of Eq.~\eqref{eq:commonfct} shows that the transport process is a
balance of drain occupation and effective average occupation, and the choice of approximation yields an overall
scaling of the corresponding emission and absorption processes. 

These expressions allow us to directly verify the thermodynamic limit for $N\to \infty$ in specific limits.

For $\bar{n} \ll N$ we have a linear scaling 
of the CGF in the system size and obtain the same result for all approximations
\bea
\lim_{N\to\infty} \frac{1}{N} {\cal C}^\alpha(\chi,t) =\mathbf{F}(\chi,t)\,. 
\eea

If we perform the thermodynamic limit and simultaneously require $\bar{n} \gg N$, we are 
always in the super-transmittance regime~\cite{vogl2011}. Scaling only the source occupation, with a finite drain occupation,
we recover the quadratic scaling of the CGF on the system size for all approximations
\bea
\lim_{n_S\to\infty} {\cal C}^\alpha(\chi,t)&=&\frac{\Gamma_D}{x_\alpha} N (N+2) \biggl[\left(e^{+i\chi}-1\right)\left(1+n_D\right)\nn 
&&+ \left(e^{-i\chi}-1\right) n_D\biggr]t\,,
\eea
where the prefactor gives a deviating scaling for Approximation 2 ($x_2=4$) but agrees with the exact solution ($x_1=x_3=6$) 
for the high thermo-bias, cf.~\cite{vogl2011}, otherwise. 
This indicates that the coherent effect of "super-transmittance"~\cite{vogl2011} is correctly found also by the 
factorization approach.

As a further benchmark for our approximations we check the limit of low thermo-bias, where for low
source occupations $n_S \ll 1$ and vanishing drain occupations $n_D = 0$ we again exactly recover previous findings (cf.~\cite{vogl2011}) for 
the master equation solution
\bea
{\cal C}(\chi,t)&=&\frac{\Gamma_S \Gamma_D}{\Gamma_S + \Gamma_D}\left(e^{+\ii \chi}-1\right) n_S N t\,,
\eea
with all approximations. 
\subsection{Scaling of stationary cumulants}\label{sec:scaling} 
%
\begin{figure*}[tb]
  \begin{center}
  \begin{minipage}[t]{0.48\textwidth}
    \includegraphics[width=\textwidth,clip=true]{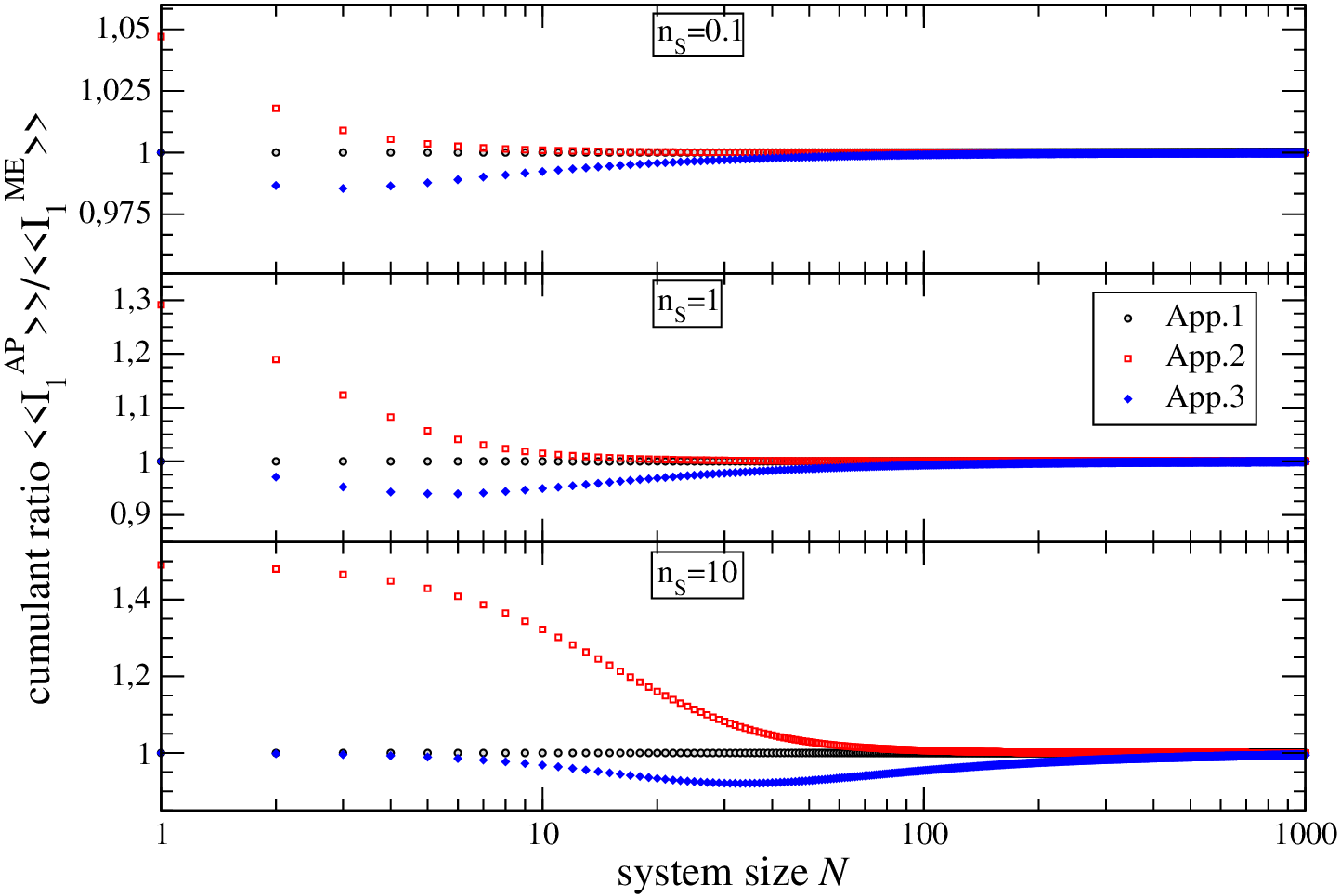} 
    \caption{\label{vergleich_C1}(Color Online)
	    Logarithmic plot of the ratio of approximated and exact master equation solution for the first steady state cumulant 
	    $\frac{\cu{I_1^{AP}}}{\cu{I_1^{ME}}}$ versus the system size $N$ for 
	    the different approximations 1 (grey filled, black circles), 2 (red squares), 3 (filled, blue diamonds), cf. Eq.~\eqref{eq:appcgf}, computed for
	    different dimensionless source occupations $n_S =0.1$ (top), $n_S=1$ (middle) and $n_S=10$ (bottom).
	    Note the different scales on the ordinate axes. 
            Other parameters: $n_D=0$.
	    }
  \end{minipage}
  \hspace{1pt}
  \begin{minipage}[t]{0.48\textwidth}
    \includegraphics[width=\textwidth,clip=true]{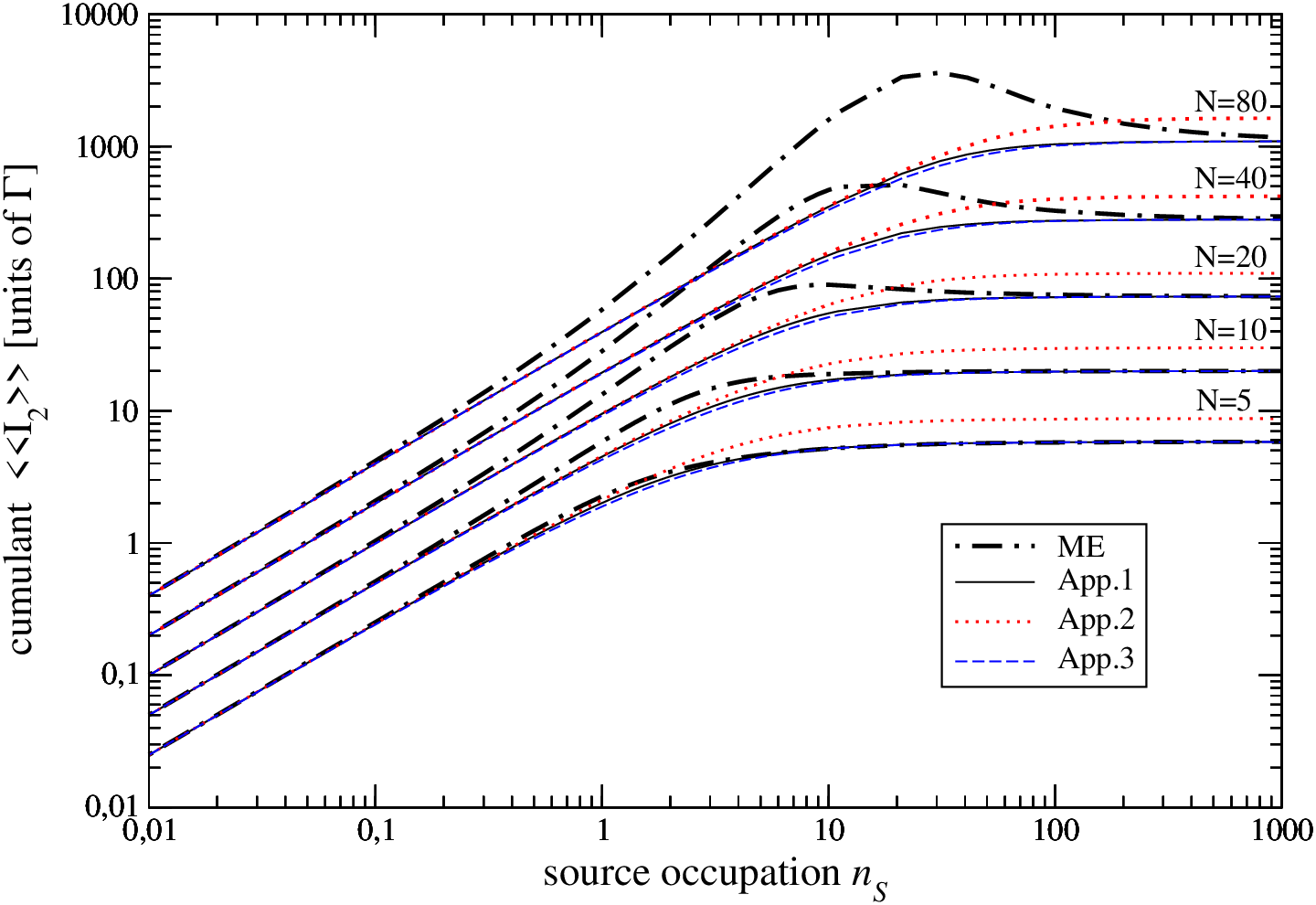}  
    \caption{\label{vergleich_C2}(Color Online)
	    Double logarithmic plot of the second steady state cumulant $\cu{I_2}$ 
	    in units of $\Gamma=\Gamma_S=\Gamma_D$ versus the dimensionless 
            source occupation $n_S$ for the numerical master equation solution 
	    (bold dot-dashed, black) and 
	    the different approximations 1 (solid, black), 2 (dotted, red), and 3 (dashed, blue), cf. Eq.~\eqref{eq:appcgf}, 
	    computed for different system size $N = 5,\,10,\,20,\,40,\,80$ (bottom to top).
	    Other parameters: $n_D=0$.
	    }
  \end{minipage}
 \end{center}
\end{figure*}
To quantify the errors which are introduced by the different approximations, we compare the corresponding long-time cumulants to 
the full master equation solution, which were derived in~\cite{vogl2011}.

First, we compare the ratio of approximate and master equation solution for the steady state of the first current cumulant 
$\frac{\cu{I_1^{AP}}}{\cu{I_1^{ME}}}$. The master equation solution is given in Eq.~\eqref{eq:n1me} and the approximate
solutions are calculated with Eq.~\eqref{eq:appcgf}. 

Interestingly, approximation 1 yields the exact analytic expression for arbitrary source or 
drain occupations $n_\alpha$, tunneling rates $\Gamma_\alpha$ and system size $N$ for the first stationary current cumulant, 
as shown by the black, solid line in all panels of Fig.~\ref{vergleich_C1}. 

For low source occupation $n_S$, cf. top panel of Fig.~\ref{vergleich_C1}, 
approximations 2 (red, dotted line) and 3 (blue, dashed line) yield 
 results comparable to the exact solution. 
For higher source occupation $n_S$ (middle and bottom panel), approximation 2 shows a deviation from the exact results
for small system size. With higher occupations it is sustained up to larger system sizes.
Approximation 3 shows a different type of deviation. The position of the maximal deviation is dependent on the 
source occupation and shifted to larger system size for higher occupations.
Additionally, as expected all approximations approach the exact solution with larger system sizes.

The approximate solutions for the second cumulant $\cu{I_2^{AP}}$ can be analytically calculated from Eq.~\eqref{eq:appcgf}.
However, for the master equation results $\cu{I_2^{ME}}$, we have to revert to numerical solutions.
Since we are interested in the validity of our factorization Eq.~\eqref{eq:factoring_assumption} for a broader range of occupations, we
compare in Fig.~\ref{vergleich_C2} the approximated second cumulants $\cu{I_2^{AP}}$ and the numerical master equation solution
$\cu{I_2^{ME}}$ for different system sizes $N=\left\{5,\,10,\,20,\,40,\,80\right\}$ over the range of source occupations 
$n_S\in\{0.01 \dots 1000\}$ (setting $n_D=0$). 

While for low source occupations all approximations show perfect agreement with the master equation solution (thick, black dot-dashed line), 
approximation 2 (red, dotted line) shows a constant deviation to the master equation solution for large source occupations. 
However, approximations 1 (thin, black solid line) and 3 (blue, dashed line) show excellent 
agreement, again. This is valid for all considered system sizes. 
For intermediate source occupations all approximations show large deviations from the master equation solution in higher cumulants.
\subsection{Quality of approximations}\label{sec:quality}
 The quality of the approximations can be understood 
by evaluating the steady state expression $\expval{J_z^\alpha}$, cf. Eq.\eqref{eq:exactJz2}, for $\alpha = 1,\,2,\,3$ in the 
corresponding regimes for $\bar{n} \to 0$ or $\bar{n} \to \infty$. In the limit of vanishing source occupation $n_S \to 0$, both 
$\expval{J_z^2}\approx \expval{J_z}^2$ and $\expval{J_z^3} \approx \expval{J_z^2}\expval{J_z}$ become exact.
In the limit of infinite source occupation $n_S \to \infty$ however, all odd expectation values $\expval{J_z^{2\alpha+1}}$ 
are vanishing, while even expectation values are scaling with powers of the system size $N$. Thus, $\expval{J_z^2}\approx \expval{J_z}^2$
is not a good approximation in this regime, while $\expval{J_z^3} \approx \expval{J_z^2}\expval{J_z}$ again gets exact, explaining
the failure of approximation 2.  

The validity of the general approximation Eq.~\eqref{eq:factoring_assumption} can be clarified by calculating
\bea
\expval{e^{\ii n \chi} J_z^\alpha} &=& \sum_{n=-\infty}^{\infty}\sum_{m=-\frac{N}{2}}^{\frac{N}{2}} \bra{n,m} e^{\ii n \chi}
J_z^\alpha \rho \ket{n,m}\\
&=&\sum_{n=-\infty}^{\infty}\sum_{m=-\frac{N}{2}}^{\frac{N}{2}} e^{\ii n \chi} \bra{m} J_z^\alpha \rho^{(n)}\ket{m}\nn
&=&\sum_{n=-\infty}^{\infty}\sum_{m=-\frac{N}{2}}^{\frac{N}{2}} (2 m)^\alpha e^{\ii n \chi} \bra{m} \rho^{(n)} \ket{m}\,,\nonumber
\eea
where $\rho^{(n)}=\bra{n}\rho\ket{n}$ is the conditioned density matrix, and in the last line the eigenvalue of $J_z^\alpha$ was 
inserted.
Obviously, the expectation value can be factored when the system is close to the ground state 
\mbox{$\bra{m}\rho^{(n)}\ket{m} \approx \delta_{m,-\frac{N}{2}} C_n$}, which is the case in the limit $\bar{n} \to 0$
(e.g. low thermobias and $n_D=0$).
Alternatively, factorization is possible when the system is equipartitioned $\bra{m}\rho^{(n)}\ket{m} \approx C_n$, 
which is fulfilled for the limit $\bar{n}\to\infty$ (e.g. large thermobias). 
This explains the good performance of the approximated solutions in the regimes of low and high weighted average occupations $\bar{n}$.

In contrast, in the intermediate region of $\bar{n}$
neither fluctuations are suppressed due to the system being close to its ground state, 
nor the supply of bosons from the source is large enough to re-pump the large spin $J_z$ to its thermalized state. 
Thus, the dynamics is governed by higher order correlations in the intermediate regime 
that are not taken into account at this level of factorization.
\section{Conclusion}\label{sec:concl}
The application of the equation of motion method to the CGF on top of the master equation allows for the 
approximate calculation of the full dynamics of the system. 
For the special case $N=1$ the method becomes exact and recovers the FCS results obtained by diagonalization of the full 
Liouvillian.
In the limits of low and high thermo-bias the master equation results of previous 
work~\cite{vogl2011} are recovered for the full long-term CGF. Furthermore, this method enables us to directly access the thermodynamic limit, 
which is not generally possible in the master equation solution. 

For the approximate solutions we showed that it is necessary to take at least the equation for $\expval{J_z^2}$ into account and 
factorize at the level of $\expval{J_z^3}$, 
to get comparable results to the master equation in the stationary limit. 

Future work should be directed to the question whether the inclusion of higher correlations of $\expval{e^{\ii n \chi} J_z^\alpha}$ 
into the equation for the CGF yields better results for higher cumulants and also at intermediate source/drain occupations.

As is evident from the case of $N=1$, one should in general carefully examine whether for system operators $A$ our main approximation 
\mbox{$\expval{e^{\ii n\chi}A} \approx \expval{e^{\ii n\chi}}\expval{A}$} is sufficient, 
or if one has to close the system of equations at higher order. However, since the zeroth order of the method allows for closure with only a few equations,
and many relevant probability distributions are close to Gaussian and thus governed by the first two cumulants~\cite{karzig2010}, this approach 
could find application in a broader community.
\section*{Acknowledgments}
We acknowledge support by the DFG via GRK 1558 (M.V.), grants SCHA 1646/2-1 (G.S.), BRA 1528/7, BRA 1528/8, SFB 910 (T.B.).
We have benefited from discussions with D. Braun.

\end{document}